\documentclass[thmsa,letterpaper,12pt]{article}
\usepackage{amssymb}
\usepackage{amsmath}
\usepackage{sw20lart}

\input tcilatex
\begin{document}

\author{Nick Laskin\thanks{%
E-mail: nlaskin@rocketmail.com}}
\title{\textbf{Fractional Classical Mechanics}\\
}
\date{TopQuark Inc.\\
Toronto, ON, M6P 2P2}
\maketitle

\begin{abstract}
Fractional classical mechanics has been introduced and developed as a
classical counterpart of the fractional quantum mechanics. Lagrange,
Hamilton and Hamilton-Jacobi frameworks have been implemented for the
fractional classical mechanics. The Lagrangian of fractional classical
mechanics has been introduced, and equation of motion has been obtained.
Fractional oscillator model has been launched and solved in 1D case. A new
equation for the period of oscillations of fractional classical oscillator
has been found. The interplay between the energy dependency of the period of
classical oscillations and the non-equidistant distribution of the energy
levels for fractional quantum oscillator has been discussed.

We discuss as well, the relationships between new equations of fractional
classical mechanics and the well-known fundamental equations of classical
mechanics.


\end{abstract}

\section{Introduction}

We introduce and elaborate fractional classical mechanics as a classical
counterpart of the fractional quantum mechanics \cite{Laskin1}-\cite{Laskin4}%
. To begin with, let us consider the equation for canonical classical
mechanics action $S_{\alpha }(p,q)$ (see, Eq.(23) in \cite{Laskin2}),

\begin{equation}
S_{\alpha }(p,q)=\int\limits_{t_{a}}^{t_{b}}d\tau (p(\tau )\overset{\cdot }{q%
}(\tau )-H_{\alpha }(p(\tau ),q(\tau ))),  \label{eq1}
\end{equation}

where the classical Hamiltonian $H_{\alpha }(p(\tau ),q(\tau ))$ arrives
from the Hamiltonian $H_{\alpha }(p,q)$ \cite{Laskin1}-\cite{Laskin4}

\begin{equation}
H_{\alpha }(p,q)=D_{\alpha }|p|^{\alpha }+V(q),\qquad 1<\alpha \leq 2,
\label{eq2}
\end{equation}

with $V(q)$ as a potential energy and $p\rightarrow p(\tau )$, $q\rightarrow
q(\tau )$, and $\{p(\tau ),q(\tau )\}$ is the particle trajectory in phase
space. The scaling factor $D_{\alpha }$ has physical dimension $[D_{\alpha
}]=\mathrm{erg}^{1-\alpha }\cdot \mathrm{cm}^{\alpha }\cdot \mathrm{sec}%
^{-\alpha }$, see, \cite{Laskin1}-\cite{Laskin2}. Since the Hamiltonian $%
H_{\alpha }(p,q)$ does not explicitly depend on the time, it represents a
conserved quantity which is in fact the total energy of fractional classical
mechanics system.

Here we consider 1D fractional classical mechanics. The 3D generalization is
straightforward and it is based on the Hamiltonian \cite{Laskin4}%
\begin{equation}
H_{\alpha }(\mathbf{p},\mathbf{q})=D_{\alpha }|\mathbf{p|}^{\alpha }+V(%
\mathbf{q}),\qquad 1<\alpha \leq 2,  \label{eq3}
\end{equation}

where $\mathbf{p}$ and $\mathbf{q}$ are 3D vectors.

The paper is organized as follows.

Fundamentals of fractional classical mechanics are covered in Sec.2, where
Lagrange, Hamilton and Hamilton-Jacobi frameworks have been implemented.

In Sec.3 the scaling analysis of fractional classical motion equations has
been developed based on the mechanical similarity. We discover and discuss 
\textit{fractional Kepler's third law }which is\textit{\ }a generalization
of the well-known Kepler's third law.

The motion equations for the fractional classical mechanics have been
integrated in Sec.4. Fractional classical oscillator model has been
introduced. We found an equation for the period of oscillations of
fractional classical oscillator. The map between the energy dependence of
the period of classical oscillations and the non-equidistant distribution of
quantum energy levels has been established.

In conclusion we outline the new developments.

\section{Fundamentals of fractional classical mechanics}

\subsection{The Lagrange approach}

The Lagrangian of fractional classical mechanics $L_{\alpha }(\overset{\cdot 
}{q},q)$ is defined as usual

\begin{equation}
L_{\alpha }(\overset{\cdot }{q},q)=p\overset{\cdot }{q}-H_{\alpha }(p,q),
\label{eq4}
\end{equation}

where the momentum $p$ is

\begin{equation}
p=\frac{\partial L_{\alpha }(\overset{\cdot }{q},q)}{\partial \overset{\cdot 
}{q}},  \label{eq5}
\end{equation}

and $H_{\alpha }(p,q)$ is given by Eq.(\ref{eq2}).

Hence, we obtain the Lagrangian of the fractional classical mechanics

\begin{equation}
L_{\alpha }(\overset{\cdot }{q},q)=\left( \frac{1}{\alpha D_{\alpha }}%
\right) ^{\frac{1}{\alpha -1}}\frac{\alpha -1}{\alpha }|\overset{\cdot }{q}%
|^{\frac{\alpha }{\alpha -1}}-V(q),\quad 1<\alpha \leq 2.  \label{eq6}
\end{equation}

For a free particle, $V(q)=0$, the Lagrangian of the fractional classical
mechanics is

\begin{equation}
L_{\alpha }^{(0)}(\overset{\cdot }{q},q)=\left( \frac{1}{\alpha D_{\alpha }}%
\right) ^{\frac{1}{\alpha -1}}\frac{\alpha -1}{\alpha }|\overset{\cdot }{q}%
|^{\frac{\alpha }{\alpha -1}},\quad 1<\alpha \leq 2.  \label{eq7}
\end{equation}

Further, the Euler-Lagrange equation of motion has the standard form

\begin{equation}
\frac{d}{dt}\frac{\partial L_{\alpha }(\overset{\cdot }{q},q)}{\partial 
\overset{\cdot }{q}}-\frac{\partial L_{\alpha }(\overset{\cdot }{q},q)}{%
\partial q}=0.  \label{eq8}
\end{equation}

By substituting Eq.(\ref{eq6}) into Eq.(\ref{eq8}) we find the motion
equation in Lagrangian form

\begin{equation}
\left( \frac{1}{\alpha D_{\alpha }}\right) ^{\frac{1}{\alpha -1}}\frac{1}{%
\alpha -1}\overset{\cdot \cdot }{q}|\overset{\cdot }{q}|^{\frac{2-\alpha }{%
\alpha -1}}+\frac{\partial V(q)}{\partial q}=0.  \label{eq9}
\end{equation}

This equation has to be accompanied by the initial conditions. We impose the
following initial conditions. At $t=0$, the initial displacement is denoted
by $q_{0}$ and the corresponding velocity is denoted by $\overset{\cdot }{q}%
_{0}$, that is,

\begin{equation}
q(t=0)=q_{0},\qquad \mathrm{and}\qquad \overset{\cdot }{q}(t=0)=\overset{%
\cdot }{q}_{0}.  \label{eq10}
\end{equation}

As one can see, Eq.(\ref{eq9}) has a non-linear kinematic term.

The new equation (\ref{eq9}) is fractional generalization of the well known
equation of motion of classical mechanics in the Lagrange form.

In the special case when $\alpha =2$, Eq.(\ref{eq9}) goes into

\begin{equation}
m\overset{\cdot \cdot }{q}+\frac{\partial V(q)}{\partial q}=0.  \label{eq11}
\end{equation}

where $m$ is a particle mass ($D_{2}=1/2m$) and the initial conditions are
given by Eq.(\ref{eq10}).

\subsection{The Hamilton approach}

To obtain the Hamilton equations of motion for the fractional classical
mechanics we apply variational principle,

\begin{equation}
\delta S_{\alpha }(p,q)=0,  \label{eq12}
\end{equation}

where the action $S_{\alpha }(p,q)$ is given by Eq.(\ref{eq1}).

Considering the momentum $p$ and coordinate $q$ as independent variables we
can write

\begin{equation}
\delta S_{\alpha }(p,q)=\int\limits_{t_{a}}^{t_{b}}d\tau (\delta p\overset{.}%
{q}+p\delta \overset{.}{q}-\frac{\partial H_{\alpha }(p,q)}{\partial p}%
\delta p-\frac{\partial H_{\alpha }(p,q)}{\partial q}\delta q)=0.
\label{eq13}
\end{equation}

Upon integration by parts of the second term $p\delta \overset{.}{q}$, the
variation $\delta S_{\alpha }$ becomes

\begin{equation*}
\delta S_{\alpha }(p,q)=\int\limits_{t_{a}}^{t_{b}}d\tau \delta p(\overset{.}%
{q}-\frac{\partial H_{\alpha }(p,q)}{\partial p})+p\delta
q|_{t_{a}}^{t_{b}}-\int\limits_{t_{a}}^{t_{b}}d\tau \delta q(\overset{.}{p}+%
\frac{\partial H_{\alpha }(p,q)}{\partial q})=0.
\end{equation*}

Since $\delta q(t_{a})=\delta q(t_{b})=0$ at the end points of the
trajectory, the term $p\delta q$ is 0. Between the end points $\delta p$ and 
$\delta q$ can take on any arbitrary value. Hence, the variation $\delta
S_{\alpha }$ can be 0 if the following conditions are satisfied

\begin{equation}
\overset{\cdot }{q}=\frac{\partial H_{\alpha }(p,q)}{\partial p},\qquad 
\overset{\cdot }{p}=-\frac{\partial H_{\alpha }(p,q)}{\partial q}.
\label{eq14}
\end{equation}

This is, of course, the canonical Hamilton equations of motion. Thus, for
the time-independent Hamiltonian given by Eq.(\ref{eq2}) we obtain the
equations

\begin{equation}
\overset{\cdot }{q}=\alpha D_{\alpha }|p|^{\alpha -1}\mathrm{sgn}p,\qquad 
\overset{\cdot }{p}=-\frac{V(q)}{\partial q},\quad \qquad 1<\alpha \leq 2,
\label{eq15}
\end{equation}

where $\mathrm{sgn}p$ is the sign function which for nonzero values of $p$
can be defined by the formula

\begin{equation*}
\mathrm{sgn}p=\frac{p}{|p|},
\end{equation*}

where $|p|$ is the absolute value of $p$.

Equations (\ref{eq15}) are the Hamilton equations for fractional classical
mechanics system.

\subsubsection{Poisson bracket approach to fractional classical mechanics}

It is well-known that Hamiltonian classical mechanics could be reformulated
in terms of the Poisson brackets. For the two arbitrary functions $u(p,q)$
and $v(p,q)$ of variables $p$ and $q$, the Poisson bracket is defined as 
\cite{Landauv1}

\begin{equation}
\{u(p,q),v(p,q)\}=\frac{\partial u}{\partial p}\frac{\partial v}{\partial q}-%
\frac{\partial u}{\partial q}\frac{\partial v}{\partial p}.  \label{eq16}
\end{equation}

The Hamilton's equations of motion have an equivalent expression in terms of
the Poisson bracket. Indeed, suppose that $\mathrm{f}(p,q,t)$ is a function
of momentum $p$, coordinate $q$ and time $t$. Then we have

\begin{equation}
\frac{d\mathrm{f}}{dt}=\frac{\partial \mathrm{f}}{\partial t}+\left( \frac{%
\partial \mathrm{f}}{\partial p}\overset{\cdot }{p}+\frac{\partial \mathrm{f}%
}{\partial q}\overset{\cdot }{q}\right) .  \label{eq17}
\end{equation}

Substituting $\overset{\cdot }{q}$ and $\overset{\cdot }{p}$ given by Eq.(%
\ref{eq14}) yields

\begin{equation}
\frac{d\mathrm{f}}{dt}=\frac{\partial \mathrm{f}}{\partial t}+\{H_{\alpha },%
\mathrm{f}\},  \label{eq18}
\end{equation}

where $\{H_{\alpha },\mathrm{f}\}$ is the Poisson bracket,

\begin{equation}
\{H_{\alpha },\mathrm{f}\}=\frac{\partial H_{\alpha }}{\partial p}\frac{%
\partial \mathrm{f}}{\partial q}-\frac{\partial H_{\alpha }}{\partial q}%
\frac{\partial \mathrm{f}}{\partial p}.  \label{eq19}
\end{equation}

with $H_{\alpha }$ given by Eq.(\ref{eq2}).

\subsection{The Hamilton-Jacobi approach}

Having the Lagrangian (\ref{eq6}), we can present classical mechanics action
given by Eq.(\ref{eq1}) as a function of coordinate $q$,

\begin{equation}
S_{\alpha }(q)=\int\limits_{t_{a}}^{t_{b}}d\tau L_{\alpha }(\overset{\cdot }{%
q},q)=\int\limits_{t_{a}}^{t_{b}}d\tau \left\{ \left( \frac{1}{\alpha
D_{\alpha }}\right) ^{\frac{1}{\alpha -1}}\frac{\alpha -1}{\alpha }|\overset{%
\cdot }{q}|^{\frac{\alpha }{\alpha -1}}-V(q)\right\} ,\quad 1<\alpha \leq 2,
\label{eq20}
\end{equation}

Let's now treat the action $S_{\alpha }(q,t_{b})$ as a function of
coordinate $q$ and the upper limit of integration, $t_{b}$. Further, we
suppose that motion goes along the actual path, that is we consider action
as a functional of actual trajectory $q(\tau )$ in coordinate space. To
evaluate the variance $\delta S_{\alpha }(q,t_{b})$ we have to compare the
values of $S_{\alpha }(q,t_{b})$ for trajectories having a common start
point at $q(t_{a})$, but passing through different end points at the $t_{b}$%
. Thus, for the variation of the action by the variation of the end point $%
\delta q(t_{b})$ of the trajectory we have

\begin{equation}
\delta S_{\alpha }(q,t_{b})=\delta \int\limits_{t_{a}}^{t_{b}}d\tau
L_{\alpha }(\overset{\cdot }{q},q)=\frac{\partial L_{\alpha }}{\partial 
\overset{\cdot }{q}}\delta
q|_{t_{a}}^{t_{b}}+\int\limits_{t_{a}}^{t_{b}}d\tau \delta q\left( \frac{%
\partial L_{\alpha }}{\partial q}-\frac{d}{dt}\frac{\partial L_{\alpha }}{%
\partial \overset{\cdot }{q}}\right)   \label{eq21}
\end{equation}

Because of Eq.(\ref{eq8}) and $\delta q(t_{a})=0$ we obtain

\begin{equation*}
\delta S_{\alpha }(q,t_{b})=p\delta q(t_{b}),
\end{equation*}

or

\begin{equation}
p=\frac{\delta S_{\alpha }(q,t_{b})}{\delta q(t_{b})}.  \label{eq22}
\end{equation}

For simplicity, letting $\delta q(t_{b})=\delta q$ and $t_{b}=t$, we can
write

\begin{equation}
\frac{dS_{\alpha }(q,t)}{dt}=\frac{\partial S_{\alpha }(q,t)}{\partial t}+%
\frac{\partial S_{\alpha }(q,t)}{\partial q}\overset{\cdot }{q}=\frac{%
\partial S_{\alpha }(q,t)}{\partial t}+p\overset{\cdot }{q}.  \label{eq23}
\end{equation}

According to the definition of the action its total time derivative along
the trajectory is $dS_{\alpha }/dt=L_{\alpha }$, and Eq.(\ref{eq23}) can be
presented as

\begin{equation}
\frac{\partial S_{\alpha }(q,t)}{\partial t}=L_{\alpha }-p\overset{\cdot }{q}%
=-H_{\alpha }(p,q),  \label{eq24}
\end{equation}

where Eq.(\ref{eq4}) has been taken into account and $p$ is given by Eq.(\ref%
{eq22}).

Therefore, we come to the equation

\begin{equation}
\frac{\partial S_{\alpha }(q,t)}{\partial t}+H_{\alpha }(\frac{\partial
S_{\alpha }(q,t)}{\partial q},q)=0.  \label{eq25}
\end{equation}

With $H_{\alpha }(p,q)$ given by Eq.(\ref{eq2}) it takes the form

\begin{equation}
\frac{\partial S_{\alpha }(q,t)}{\partial t}+D_{\alpha }|\frac{\partial
S_{\alpha }(q,t)}{\partial q}|^{\alpha }+V(q)=0,\qquad 1<\alpha \leq 2,
\label{eq26}
\end{equation}

where $S_{\alpha }$ is called the Hamilton's principal function.

For time-independent Hamiltonian the variables $q$ and $t$ in this equation
can be separated. That is, we can search for solution of Eq.(\ref{eq26}) in
the form

\begin{equation}
S_{\alpha }(q,t,E)=S_{\alpha }^{(0)}(q,E)-Et,  \label{eq27}
\end{equation}

where $S_{\alpha }^{(0)}(q,E)$ is time-independent Hamilton's principal
function and $E$, is the constant of integration, which has been identified
with the total energy. Substituting Eq.(\ref{eq27}) into Eq.(\ref{eq26})
yields

\begin{equation}
D_{\alpha }|\frac{\partial S_{\alpha }^{(0)}(q,E)}{\partial q}|^{\alpha
}+V(q)=E,\qquad 1<\alpha \leq 2.  \label{eq28}
\end{equation}

Equations (\ref{eq26}) and (\ref{eq28}) are the Hamilton-Jacobi equations of
fractional classical mechanics.

As an example, let us consider a free particle, $V(q)=0$. The
Hamilton-Jacobi equation Eq.(\ref{eq26}) for a free particle is

\begin{equation}
\frac{\partial S_{\alpha }(q,t)}{\partial t}+D_{\alpha }|\frac{\partial
S_{\alpha }(q,t)}{\partial q}|^{\alpha }=0,\qquad 1<\alpha \leq 2.
\label{eq29}
\end{equation}

Therefore, from Eqs.(\ref{eq27}) and (\ref{eq28}) we obtain a solution to
the Hamilton-Jacobi equation Eq.(\ref{eq29}),

\begin{equation}
S_{\alpha }(q,t,E)=(\frac{E}{D_{\alpha }})^{1/\alpha }q-Et.  \label{eq30}
\end{equation}

Follow the Hamilton-Jacobi fundamentals (see, for instance, \cite{Landauv1})
we differentiate Eq.(\ref{eq30}) over the energy $E$ and put the derivative
equal to a new constant $\delta $,

\begin{equation}
\delta =\frac{\partial S_{\alpha }(q,t,E)}{\partial E}=\frac{1}{\alpha
D_{\alpha }}(\frac{E}{D_{\alpha }})^{\frac{1}{\alpha }-1}q-t,  \label{eq31}
\end{equation}

which yields

\begin{equation}
q=\alpha D_{\alpha }(\frac{E}{D_{\alpha }})^{1-\frac{1}{\alpha }}(t+\delta ),
\label{eq32}
\end{equation}

and

\begin{equation}
p=\frac{\partial S_{\alpha }(q,t,E)}{\partial q}=(\frac{E}{D_{\alpha }}%
)^{1/\alpha },  \label{eq33}
\end{equation}

where $E>0$ and $1<\alpha \leq 2$.

The equations (\ref{eq32}) and (\ref{eq33}) define the trajectory of a free
particle in fractional classical mechanics framework.

In limit case when $\alpha =2$ and $D_{2}=1/2m$, Eqs.(\ref{eq9}), (\ref{eq15}%
), (\ref{eq26}) and (\ref{eq28}) are transformed into the well-known
Hamilton, Lagrange and Hamilton-Jacobi equations of classical mechanics for
a particle with mass $m$ moving in the potential field $V(q)$.

\section{Mechanical similarity}

First of all, we intend to study the general properties of 1D fractional
classical motion without integrating motion equations (\ref{eq9}), (\ref%
{eq15}), and (\ref{eq28}).

When the potential energy $V(q)$ is a homogeneous function of coordinate $q$%
, it is possible to find some general similarity relationships.

Let us carry out a transformation in which the coordinates are changed by a
factor $\rho $ and the time by a factor $\tau $: $q\rightarrow q^{\prime
}=\rho q$, $t\rightarrow t^{\prime }=\tau t$. Then all velocities $\overset{%
\cdot }{q}=dq/dt$ are changed by a factor $\rho /\tau $, and the kinetic
energy by a factor $\left( \rho /\tau \right) ^{\alpha /(\alpha -1)}$. If $%
V(q)$ is a homogeneous function of degree $\beta $ then it satisfies,

\begin{equation}
V(\rho q)=\rho ^{\beta }V(q).  \label{eq34}
\end{equation}

It is obvious that if $\rho $ and $\tau $ are such that $\left( \rho /\tau
\right) ^{\alpha /(\alpha -1)}=\rho ^{\beta }$ , i.e. $\tau =\rho ^{1-\beta +%
\frac{\beta }{\alpha }}$, then the transformation leaves the motion equation
unaltered. This invariance is called mechanical similarity.

A change of all the coordinates of the particles by the same scale factor
corresponds to replacement of the classical mechanical trajectories of the
particles by other trajectories, geometrically similar but different in
size. Thus, we conclude that, if the potential energy of the system is a
homogeneous function of degree $\beta $ in Cartesian coordinates, the
fractional equations of motion permit a series of geometrically similar
trajectories, and the times of the motion between corresponding points are
in the ratio

\begin{equation}
\frac{t^{\prime }}{t}=\left( \frac{l^{\prime }}{l}\right) ^{1-\beta +\frac{%
\beta }{\alpha }},  \label{eq35}
\end{equation}

where $\frac{l^{\prime }}{l}$ is the ratio of linear scales of the two
paths. Not only the times but also any other mechanical quantities are in a
ratio which is a power of $\frac{l^{\prime }}{l}$. For example, the
velocities and energies follow the scaling laws

\begin{equation}
\frac{\overset{\cdot }{q}^{\prime }}{\overset{\cdot }{q}}=\left( \frac{%
l^{\prime }}{l}\right) ^{\beta -\frac{\beta }{\alpha }},\qquad \qquad \frac{%
E^{\prime }}{E}=\left( \frac{l^{\prime }}{l}\right) ^{\beta },  \label{eq36}
\end{equation}

and

\begin{equation}
\frac{t^{\prime }}{t}=\left( \frac{E^{\prime }}{E}\right) ^{\frac{1}{\alpha }%
+\frac{1}{\beta }-1}.  \label{eq37}
\end{equation}

The following are some examples of the foregoing.

It follows from Eq.(\ref{eq37}) that at $\frac{1}{\alpha }+\frac{1}{\beta }%
=1 $ the period of oscillations does not depend on energy of oscillator. The
condition $\frac{1}{\alpha }+\frac{1}{\beta }=1$ with $1<\alpha \leq 2$, $%
1<\beta \leq 2$ brings $\alpha =2$ and $\beta =2$ only. It means, for
instance, that considering the fractional classical oscillator model with
Hamiltonian (\ref{eq41}), we can conclude that the standard classical
harmonic oscillator only,

\begin{equation}
H_{2}(p,q)=p^{2}/2m+g^{2}q^{2},  \label{eq38}
\end{equation}

has the period of oscillations which does not depend on energy of oscillator.

In the uniform field of force, the potential energy is a linear function of
the coordinates, i.e. $\beta =1$. From Eq.(\ref{eq37}) we have $\frac{%
t^{\prime }}{t}=\left( \frac{l^{\prime }}{l}\right) ^{1/\alpha }$.
Therefore, the time of motion in a uniform field ($\beta =1)$ is as the $%
\alpha $-root of the initial altitude.

If the potential energy is inversely proportional to the distance apart,
i.e. it is a homogeneous function of degree $\beta =-1$, then Eq.(\ref{eq35}%
) becomes

\begin{equation}
\frac{t^{\prime }}{t}=\left( \frac{l^{\prime }}{l}\right) ^{2-\frac{1}{%
\alpha }}.  \label{eq39}
\end{equation}

For instance, regarding the problem of orbital motion in 3D fractional
classical mechanics, we can state that the orbital period to the power $%
\alpha $ is proportional to the power $2\alpha -1$ of its orbit scale. This
statement is in fact a generalization of the well-known Kepler's third law.
We call Eq.(\ref{eq39}) as \textit{fractional Kepler's third law}. In
special case $\alpha =2$, Eq.(\ref{eq39}) turns into the well-known Kepler's
third law \cite{Landauv1}.

In general, for negative $\beta $, $\beta =-\gamma $, in the space of the
fractional Hamiltonians $H_{\alpha ,-\gamma }$ at

\begin{equation}
1+\gamma -\frac{\gamma }{\alpha }=\frac{3}{2},  \label{eq40}
\end{equation}

there exists the subset of fractional dynamic systems, orbital motion of
which follows the well-known Kepler's third law, that is, the square of the
orbital period is proportional to the cube of the orbit size.

\section{Integration of the motion equations}

\subsection{Fractional classical 1D oscillator}

\subsubsection{The Lagrange approach}

Fractional quantum oscillator model has been introduced in \cite{Laskin1}.
Fractional classical oscillator can be considered as a classical counterpart
of the fractional quantum oscillator model.

We introduce fractional classical 1D oscillator as a mechanical system with
the Hamiltonian

\begin{equation}
H(p,q)=D_{\alpha }|p|^{\alpha }+g^{2}|q|^{\beta },  \label{eq41}
\end{equation}

where $g$ is a constant with physical dimension $[g]=\mathrm{erg}^{1/2}\cdot 
\mathrm{cm}^{-\beta /2}$ and $\alpha $ and $\beta $ are parameters, $%
1<\alpha \leq 2$, $1<\beta \leq 2$.

It follows from Eq.(\ref{eq6}) that the Lagrangian of the fractional
classical 1D oscillator is

\begin{equation}
L(\overset{\cdot }{q},q)=\left( \frac{1}{\alpha D_{\alpha }}\right) ^{\frac{1%
}{\alpha -1}}\frac{\alpha -1}{\alpha }|\overset{\cdot }{q}|^{\frac{\alpha }{%
\alpha -1}}-g^{2}|q|^{\beta },\quad 1<\alpha \leq 2,\quad 1<\beta \leq 2.
\label{eq42}
\end{equation}

Hence, the motion equation of fractional classical 1D oscillator is

\begin{equation}
\left( \frac{1}{\alpha D_{\alpha }}\right) ^{\frac{1}{\alpha -1}}\frac{1}{%
\alpha -1}\overset{\cdot \cdot }{q}|\overset{\cdot }{q}|^{\frac{2-\alpha }{%
\alpha -1}}+\beta g^{2}|q|^{\beta -1}\mathrm{sgn}q=0.  \label{eq43}
\end{equation}

This is a new non-linear classical mechanics equation of motion. When $%
\alpha =2$ and $\beta =2$ it goes into the well-known linear equation of
motion for classical mechanics oscillator which has the form,

\begin{equation}
m\overset{\cdot \cdot }{q}+2g^{2}q=0.  \label{eq44}
\end{equation}

where $m$ is a particle mass ($D_{2}=1/2m$).

To integrate Eq.(\ref{eq43}) we start from the law of energy conservation.
For the Lagrangian Eq.(\ref{eq42}) we have

\begin{equation}
D_{\alpha }\left( \frac{1}{\alpha D_{\alpha }}\right) ^{\frac{\alpha }{%
\alpha -1}}|\overset{\cdot }{q}|^{\frac{\alpha }{\alpha -1}}+g^{2}|q|^{\beta
}=E,  \label{eq45}
\end{equation}

where $E$ is the total energy. Thus, we have

\begin{equation}
|\overset{\cdot }{q}|=\alpha D_{\alpha }^{1/\alpha }(E-g^{2}|q|^{\beta })^{%
\frac{\alpha -1}{\alpha }},  \label{eq46}
\end{equation}

which is a first-order differential equation, and it can be integrated.

Since the kinetic energy is positive, the total energy $E$ always exceeds
the potential energy, that is $E>g^{2}|q|^{\beta }$. The points where the
potential energy equals the total energy $g^{2}|q|^{\beta }=E$ are turning
points of classical trajectory. The 1D motion bounded by two turning points
is oscillatory, the particle moves repeatedly between those two points.

Substituting $q=(E/g^{2})^{1/\beta }y$ into Eq.(\ref{eq46}) yields

\begin{equation}
|\overset{\cdot }{y}|=\alpha D_{\alpha }^{1/\alpha }g^{2/\beta }E^{1-(\frac{1%
}{\alpha }+\frac{1}{\beta })}(1-|y|^{\beta })^{1-\frac{1}{\alpha }}.
\label{eq47}
\end{equation}

Hence, the period $T(\alpha ,\beta )$ of oscillations is

\begin{equation*}
T(\alpha ,\beta )=4\frac{E^{(\frac{1}{\alpha }+\frac{1}{\beta })-1}}{\alpha
D_{\alpha }^{1/\alpha }g^{2/\beta }}\int\limits_{0}^{1}\frac{dy}{(1-y^{\beta
})^{1-\frac{1}{\alpha }}}.
\end{equation*}

By substituting $z=y^{\beta }$ we rewrite the last equation in the form 
\begin{equation}
T(\alpha ,\beta )=4\frac{E^{(\frac{1}{\alpha }+\frac{1}{\beta })-1}}{\alpha
\beta D_{\alpha }^{1/\alpha }g^{2/\beta }}\int\limits_{0}^{1}dzz^{\frac{1}{%
\beta }-1}(1-z)^{\frac{1}{\alpha }-1}.  \label{eq48}
\end{equation}

With the help of the $B$-function definition \cite{DLMFB}

\begin{equation*}
B(\frac{1}{\beta },\frac{1}{\alpha })=\int\limits_{0}^{1}dzz^{\frac{1}{\beta 
}-1}(1-z)^{\frac{1}{\alpha }-1},
\end{equation*}

we finally find for the period of oscillations of fractional classical 1D
oscillator

\begin{equation}
T(\alpha ,\beta )=4\frac{E^{(\frac{1}{\alpha }+\frac{1}{\beta })-1}}{\alpha
\beta D_{\alpha }^{1/\alpha }g^{2/\beta }}B(\frac{1}{\beta },\frac{1}{\alpha 
}).  \label{eq49}
\end{equation}

This new equation shows that the period depends on the energy of fractional
classical 1D oscillator. The dependency on energy is in agreement with the
scaling law given by Eq.(\ref{eq37}).

It follows from Eq.(\ref{eq49}) that period $T(\alpha ,\beta )$ doesn't
depend on the energy of oscillator when $(\frac{1}{\alpha }+\frac{1}{\beta }%
)=1$. If $1<\alpha \leq 2$ and $1<\beta \leq 2$ then condition $(\frac{1}{%
\alpha }+\frac{1}{\beta })=1$ gives us that $\alpha =2$ and $\beta =2$.
Hence, we come to the standard classical mechanics harmonic oscillator with
the Hamiltonian given by Eq.(\ref{eq38}) and the energy independent
oscillation period, $T(2,2)=\pi \sqrt{2m}/g$.

Table 1 shows that the energy dependency of the period of fractional
classical oscillator is a classical counterpart of the fractional quantum
mechanics statement on non-equidistant energy levels of fractional quantum
oscillator. For classical mechanics, independence on energy of the period of
classical oscillator is a classical counterpart of the quantum mechanics
statement on equidistant energy levels of quantum oscillator.

\begin{equation*}
\end{equation*}

\begin{tabular}{|c|c|c|}
\hline
& Period of oscillations $T$ & Energy levels $\mathcal{E}_{n}$ \\ \hline
$\QATOP{{\Large 1<\alpha <2}}{{\Large 1<\beta <2}}$ & $\QATOP{\text{%
Fractional classical mechanics}}{{\Large 4}\frac{E^{(\frac{1}{\alpha }+\frac{%
1}{\beta })-1}}{\alpha \beta D_{\alpha }^{1/\alpha }g^{1/\beta }}{\Large B(}%
\frac{1}{\beta }{\Large ,}\frac{1}{\alpha }{\Large )}}$ & $\QATOP{\text{%
Fractional quantum mechanics}}{\left( \frac{\pi \hbar \beta D_{\alpha
}^{1/\alpha }g^{2/\beta }}{2B(\frac{1}{\beta },\frac{1}{\alpha }+1)}\right)
^{\frac{\alpha \beta }{\alpha +\beta }}(n+\frac{1}{2})^{\frac{\alpha \beta }{%
\alpha +\beta }}}$ \\ \hline
$\QATOP{{\Large \alpha =2}}{{\Large \beta =2}}$ & $\QATOP{\text{Classical
mechanics}}{\pi \sqrt{2m}/g}$ & $\QATOP{\text{Quantum mechanics}}{\hbar 
\sqrt{\frac{2}{m}}g(n+\frac{1}{2})}$ \\ \hline
\end{tabular}

Table 1. \textit{Energy dependence/independence of the period of classical
oscillator vs non-equidistant/equidistant energy levels of quantum oscillator%
}\footnote{%
The period of oscillations $T$ and frequency of oscillations $\omega $ are
related by $T=2\pi /\omega .$ Hence, the classical mechanics frequency is $%
\omega =$ $\sqrt{\frac{2}{m}}g$, and in terms of the frequency, the quantum
mechanic energy levels are $\mathcal{E}_{n}=\hbar \omega (n+\frac{1}{2})$.
\par
The energy levels equation for the fractional quantum oscillator has been
taken from \cite{Laskin4}.
\par
{}}.

\subsubsection{The Hamilton approach}

For the fractional 1D oscillator the Hamilton equations of motion in
accordance with Eqs.(\ref{eq14}) and (\ref{eq41}) are

\begin{equation}
\overset{\cdot }{q}=\frac{\partial H_{\alpha ,\beta }}{\partial p}=\alpha
D_{\alpha }|p|^{\alpha -1}\mathrm{sgn}p,  \label{eq50}
\end{equation}

\begin{equation}
\overset{\cdot }{p}=-\frac{\partial H_{\alpha ,\beta }}{\partial q}=-\beta
g^{2}|q|^{\beta -1}\mathrm{sgn}q,  \label{eq51}
\end{equation}

where $1<\alpha \leq 2,\quad 1<\beta \leq 2$.

Further, Hamilton equation (\ref{eq50}) leads to

\begin{equation}
|p|=\left( \frac{1}{\alpha D_{\alpha }}\right) ^{\frac{1}{\alpha -1}}|%
\overset{\cdot }{q}|^{\frac{1}{\alpha -1}},  \label{eq52}
\end{equation}

Substituting Eq.(\ref{eq52}) into Eq.(\ref{eq41}) yields exactly Eq.(\ref%
{eq45}). Therefore, downstream integration can be done by the same way as it
was done above in the framework of the Lagrange approach.

\subsubsection{The Hamilton-Jacobi approach}

For fractional classical oscillator with the Hamilton function defined by
Eq.(\ref{eq41}) a complete integral of the Hamilton-Jacobi equation (\ref%
{eq27}) is

\begin{equation}
S(q,t,E)=\dint dq(\frac{1}{D_{\alpha }}(E-g^{2}|q|^{\beta }))^{1/\alpha }-Et,
\label{eq53}
\end{equation}

where the integration constant $E$ can be identified with the total energy
of the fractional classical oscillator.

It is known that in standard classical mechanics \cite{Landauv1}, $\alpha =2$%
, the integral in Eq.(\ref{eq53}) is considered as an indefinite integral.
At this point, to move forward, we introduce the ansatz to treat the
integral in Eq.(\ref{eq53}) as

\begin{equation}
S(q,t,E)=\dint\limits_{0}^{q}dq(\frac{1}{D_{\alpha }}(E-g^{2}|q|^{\beta
}))^{1/\alpha }-Et.  \label{eq54}
\end{equation}

Following the Hamilton-Jacobi fundamentals (see, for instance, \cite%
{Landauv1}) we differentiate Eq.(\ref{eq53}) over the energy $E$ and put the
derivative equal to a new constant $\delta $,

\begin{equation}
\frac{\partial S(q,t,E)}{\partial E}=\frac{1}{\alpha D_{\alpha }^{1/\alpha }}%
\dint\limits_{0}^{q}dq^{\prime }(E-g^{2}|q^{\prime }|^{\beta })^{(1-\alpha
)/\alpha }-t=\delta .  \label{eq55}
\end{equation}

Substituting integration variable $q^{\prime }$ with new variable $z$, $%
q^{\prime }=(E/g^{2})^{1/\beta }z^{1/\beta }$ yields

\begin{equation}
t+\delta =\frac{E^{(\frac{1}{\alpha }+\frac{1}{\beta })-1}}{\alpha \beta
D_{\alpha }^{1/\alpha }g^{2/\beta }}\dint\limits_{0}^{q^{\beta }g^{2}/E}dzz^{%
\frac{1}{\beta }-1}(1-z)^{\frac{1}{\alpha }-1}.  \label{eq56}
\end{equation}

The integral in the above equation can be expressed as the incomplete Beta
function\footnote{%
The incomplete Beta function is defined as \cite{DLMFL}
\par
{}
\par
\begin{equation}
B_{x}(\mu ,\nu )=\dint\limits_{0}^{x}dyy^{\mu -1}(1-y)^{\nu -1},
\label{eq57}
\end{equation}%
} and we obtain

\begin{equation}
t+\delta =\frac{E^{(\frac{1}{\alpha }+\frac{1}{\beta })-1}}{\alpha \beta
D_{\alpha }^{1/\alpha }g^{2/\beta }}B_{q^{\beta }g^{2}/E}(\frac{1}{\beta },%
\frac{1}{\alpha }).  \label{eq58}
\end{equation}

This equation is the solution of the fractional 1D oscillator as function $%
t(q)$ in terms of the incomplete Beta function.

The incomplete Beta function $B_{x}(\mu ,\nu )$ has the hypergeometric
representation

\begin{equation}
B_{x}(\mu ,\nu )=\frac{x^{\mu }}{\mu }F(\mu ,1-\nu ;\mu +1;x),  \label{eq59}
\end{equation}

where $F(\mu ,1-\nu ;\mu +1;x)$ is the hypergeometric function \cite{DLMFF}.

Now we can write Eq.(\ref{eq58}) in the form

\begin{equation}
t+\delta =\frac{E^{\frac{1}{\alpha }-1}}{\alpha D_{\alpha }^{1/\alpha }}qF(%
\frac{1}{\beta },1-\frac{1}{\alpha };\frac{1}{\beta }+1;q^{\beta }g^{2}/E).
\label{eq60}
\end{equation}

Thus, in terms of the hypergeometric function we found solution of the
fractional 1D oscillator as function $t(q)$.

Let us show that the ansatz given by Eq.(\ref{eq54}) and based on it
solution given by Eq.(\ref{eq60}) allow us to reproduce the well-known
solution to standard classical harmonic oscillator, $\alpha =2$ and $\ \beta
=2$. Indeed, at $\alpha =2$ and $\ \beta =2$ we have

\begin{equation}
\omega (t+\delta )=xF(\frac{1}{2},\frac{1}{2};\frac{3}{2};x^{2}),
\label{eq61}
\end{equation}

where a new variable $x$ has been introduced as,

\begin{equation}
x=g\sqrt{1/E}q,  \label{eq62}
\end{equation}

and $\omega =\sqrt{2/m}g$ is the frequency of classical 1D oscillator. Since 
$xF(\frac{1}{2},\frac{1}{2};\frac{3}{2};x^{2})=\arcsin x$, it follows from
Eq.(\ref{eq61}) that

\begin{equation*}
x(t)=\sin \omega (t+\delta ).
\end{equation*}

By restoring the original dynamic variable $q(t)$ from Eq.(\ref{eq62}) we
obtain,

\begin{equation}
q(t)=\sqrt{\frac{2E}{m\omega ^{2}}}\sin \omega (t+\delta ),  \label{eq63}
\end{equation}

which is the solution of the Hamiltron-Jacobi equation for standard 1D
harmonic oscillator with the Hamiltonian given by Eq.(\ref{eq38}).

Thus, we proved that the ansatz given by Eq.(\ref{eq54}) gives us the
well-known solution to standard classical mechanics harmonic oscillator.

\section{Conclusion}

Fractional classical mechanics has been introduced as a classical
counterpart of fractional quantum mechanics. The Lagrange, Hamilton and
Hamilton-Jacobi frameworks have been developed for fractional classical
mechanics. Scaling analysis of fractional classical motion equations has
been implemented based on the mechanical similarity. We discover and discuss 
\textit{fractional Kepler's third law }which is\textit{\ }a generalization
of the well-known Kepler's third law.

Fractional classical oscillator model has been introduced and motion
equations for the fractional classical oscillator have been integrated. We
found an equation for the period of oscillations of fractional classical
oscillator. The map between the energy dependence of the period of classical
oscillations and the non-equidistant distribution of the energy levels for
fractional quantum oscillator has been established.

In the case when $\alpha =2$, all new developments are turned into the
well-known results of the classical mechanics.

\end{document}